\documentclass[superscriptaddress,amsmath,amssymb,aps,prc,singlecolumn]{revtex4-2}

\usepackage[T1]{fontenc}

\usepackage{graphicx}% Include figure files
\usepackage[hidelinks]{hyperref}% add hypertext capabilities

\usepackage{siunitx} % SI unit formatting
\usepackage[version=4]{mhchem} % chemical notations
\usepackage{tabularx} % more control over tables
\usepackage{hyperref}

\usepackage{float}
\usepackage{subfigure}
\usepackage{orcidlink}

\begin{document}

\title{Refining the nuclear mass surface with the mass of $^{103}$Sn}% Force line breaks with \\

\author{L.~Nies\orcidlink{0000-0003-2448-3775}}
\email{Lukas.Nies@cern.ch}
\affiliation{European Organization for Nuclear Research (CERN), 1211 Geneva 23, Switzerland}
\affiliation{Institut für Physik, Universität Greifswald, 17487 Greifswald, Germany}

\author{D.~Atanasov\orcidlink{0000-0002-0491-4710}}
\altaffiliation{Present address: Belgian Nuclear Research Centre SCK CEN, Boeretang 200, 2400 Mol, Belgium}
\affiliation{LP2i Bordeaux, UMR5797, Université de Bordeaux, CNRS, France}

\author{M.~Athanasakis-Kaklamanakis\orcidlink{0000-0003-0336-5980}}
\altaffiliation{Present address: Centre for Cold Matter, Imperial College London, London SW7 2AZ, United Kingdom}
\affiliation{European Organization for Nuclear Research (CERN), 1211 Geneva 23, Switzerland}
\affiliation{KU Leuven, Instituut voor Kern- en Stralingsfysica, B-3001 Leuven, Belgium}
% ORCID: 0000-0003-0336-5980

\author{M.~Au\orcidlink{0000-0002-8358-7235}}
\affiliation{European Organization for Nuclear Research (CERN), 1211 Geneva 23, Switzerland}
\affiliation{Johannes Gutenberg-Universit\"at Mainz, 55099 Mainz, Germany}

\author{C.~Bernerd\orcidlink{0000-0002-2183-9695}}
\affiliation{European Organization for Nuclear Research (CERN), 1211 Geneva 23, Switzerland}

\author{K.~Blaum\orcidlink{0000-0003-4468-9316}}
\affiliation{Max-Planck-Institut für Kernphysik, 69117 Heidelberg, Germany}
% ORCID: 0000-0003-4468-9316

\author{K.~Chrysalidis\orcidlink{0000-0003-2908-8424}}
\affiliation{European Organization for Nuclear Research (CERN), 1211 Geneva 23, Switzerland}

\author{P.~Fischer\orcidlink{0000-0002-1985-313X}}
\affiliation{Institut für Physik, Universität Greifswald, 17487 Greifswald, Germany}
% ORCID: 0000-0002-1985-313X

\author{R.~Heinke\orcidlink{0000-0001-6053-1346}}
\affiliation{European Organization for Nuclear Research (CERN), 1211 Geneva 23, Switzerland}
% ORCID: 0000-0001-6053-1346

\author{C.~Klink\orcidlink{0000-0002-9977-0106}}
\affiliation{European Organization for Nuclear Research (CERN), 1211 Geneva 23, Switzerland}
\affiliation{Institut für Kernphysik, Technische Universität Darmstadt, 64289 Darmstadt, Germany}

\author{D.~Lange\orcidlink{0000-0002-4559-6739}}
\affiliation{Max-Planck-Institut für Kernphysik, 69117 Heidelberg, Germany}
% ORCID: 0000-0002-4559-6739

\author{D.~Lunney\orcidlink{0000-0002-3227-305X}}
\affiliation{Université Paris-Saclay, CNRS/IN2P3, IJCLab, 91405 Orsay, France}

\author{V.~Manea\orcidlink{0000-0003-2065-5517}}
\affiliation{Université Paris-Saclay, CNRS/IN2P3, IJCLab, 91405 Orsay, France}

\author{B.~A.~Marsh\orcidlink{0000-0002-6184-5619}}
\altaffiliation{deceased}
\affiliation{European Organization for Nuclear Research (CERN), 1211 Geneva 23, Switzerland}

\author{M.~M\"uller\orcidlink{0000-0002-7281-9002}}
\affiliation{Max-Planck-Institut für Kernphysik, 69117 Heidelberg, Germany}

\author{M.~Mougeot\orcidlink{0000-0003-1372-1205}}
\affiliation{University of Jyvaskyla, Department of Physics, Accelerator laboratory, P.O. Box 35(YFL) FI-40014 University of Jyvaskyla, Finland}
\affiliation{Max-Planck-Institut für Kernphysik, 69117 Heidelberg, Germany}

\author{S.~Naimi\orcidlink{0000-0003-4210-8741}}
\affiliation{Université Paris-Saclay, CNRS/IN2P3, IJCLab, 91405 Orsay, France}

\author{Ch.~Schweiger\orcidlink{0000-0002-7039-1989}}
\affiliation{Max-Planck-Institut für Kernphysik, 69117 Heidelberg, Germany}
% ORCID 0000-0002-7039-1989

\author{L.~Schweikhard\orcidlink{0009-0002-8272-0388}}
\affiliation{Institut für Physik, Universität Greifswald, 17487 Greifswald, Germany}

\author{F.~Wienholtz\orcidlink{0000-0002-4367-7420}}
\affiliation{Institut für Kernphysik, Technische Universität Darmstadt, 64289 Darmstadt, Germany}

\author{the ISOLDE Collaboration}

% \collaboration{ISOLTRAP Collaboration}%\noaffiliation

\date{\today}% It is always \today, today,

\begin{abstract}
Mass measurements with the ISOLTRAP mass spectrometer at CERN-ISOLDE improve mass uncertainties of neutron-deficient tin isotopes towards doubly-magic $^{100}$Sn.
The mass uncertainty of $^{103}$Sn was reduced by a factor of 4, and the new value for the mass excess of $\SI{-67104\pm 18}{\kilo\electronvolt}$ is compared with nuclear \textit{ab initio} and density functional theory calculations.
Based on these results and local trends in the mass surface, the masses of $^{101,103}$Sn, as determined through their $Q_{\textrm{EC}}$ values, were found to be inconsistent with the new results.
From our measurement for $^{103}$Sn, we extrapolate the mass excess of $^{101}$Sn to $\SI{-60005\pm 300}{\kilo\electronvolt}$, which is significantly more bound than previously suggested. 
By correcting the mass values for $^{101,103}$Sn, we also adjust the values of $^{104}$Sb, $^{105,107}$Te, $^{108}$I, $^{109,111}$Xe, and $^{112}$Cs near the proton drip line which are connected through their $\alpha$- and proton $Q$-values.
The results show an overall smoothening of the mass surface, suggesting the absence of deformation energy above the ${N=50}$ shell closure.

\end{abstract}

\maketitle

\section{\label{sec:Introduction}Introduction}

The heaviest self-conjugate ($N=Z$) nucleus bound against proton emission, $^{100}$Sn, is one of the most interesting isotopes for nuclear research~\cite{FAESTERMANN201385}.
Its experimental investigation is important for testing our understanding of nuclear forces, especially since it is theorized to be a doubly magic nucleus with ${N=Z=50}$ nucleons.
First experimentally observed in the 1990s at GANIL, France~\cite{LEWITOWICZ199420}, and at GSI, Germany~\cite{Schneider1994}, it has the smallest $\textrm{log}(ft)$ value of any super-allowed Gamow-Teller $\beta$-decaying nuclei~\cite{Hinke2012_fullauthors}, and it marks the endpoint of a secluded island of alpha-emitters~\cite{Auranen_2018_Alpha_decay_to_100Sn, Darby_2010_alpha_decay_109Xe} and of the astrophysical $rp$-process~\cite{Schatz_2001_end_of_rp, Elomaa_2009_tin, 2023_Zhou}.

Due to its small production cross-section through nuclear reactions~\cite{2022_Qu_100Sn_production}, experimental knowledge on $^{100}$Sn and its direct neighbors is, however, still sparse.
Many investigations in this neutron-deficient area of the nuclear chart have been performed at facilities such as GSI, Germany, and RIKEN, Japan, using heavy primary beams such as $^{112}$In or $^{124}$Xe to produce rare isotopes through fragmentation processes~\cite{Hinke2012_fullauthors, Park_2020_Cd_In_Spectroscopy, 2019_Lubos_100Sn_gamowteller_fullaothors, Hornung_2020_97Ag_observation, MOLLAEBRAHIMI_2023_Cd_Rh} or fusion-evaporation reactions using the IGISOL technique \cite{2008_Kankainen_94Ag, Ge_2024_Silver_masses} in Jyvaskyla, Finland.
Using the Isotope Separator Online (ISOL) technique with $\SI{1.4}{\giga\electronvolt}$ protons accelerated onto a thick target at CERN-ISOLDE in Switzerland~\cite{2017_Catherall_ISOLDE}, masses of indium isotopes ($Z=49$) and their long-lived excited states were measured including $^{99}$In, one proton below $^{100}$Sn~\cite{2021_Mougeot, 2023_Nies_In}, while laser spectroscopy has been performed on tin isotopes down to $^{104}$Sn~\cite{2022_CRIS_in_prep} and Coulomb excitation down to $^{106}$Sn~\cite{2008_Ekstrom_Sn_coulex}.

\begin{figure*}[t!]
	\centering
	\includegraphics[width=1\textwidth]{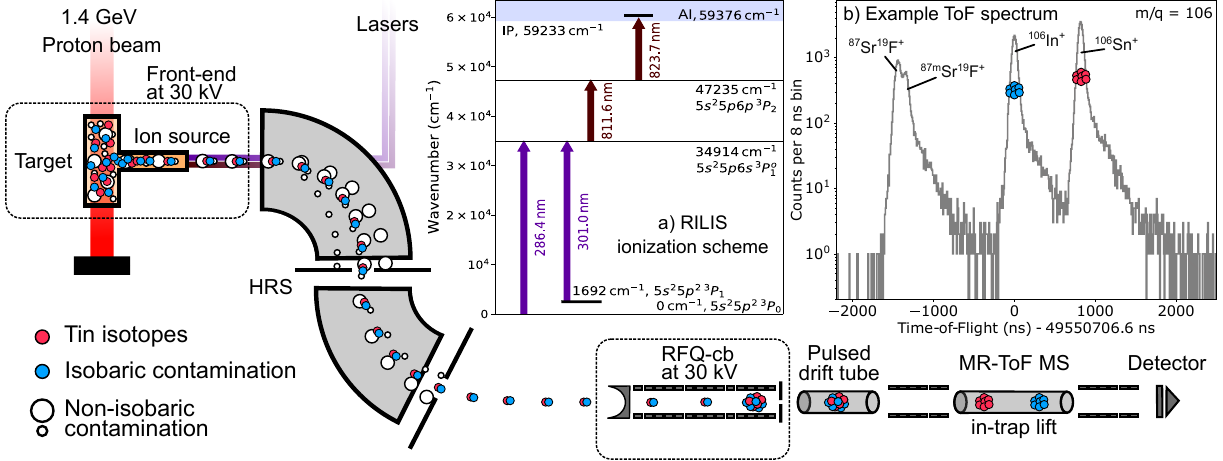}
	\caption[]{Schematic of the apparatus. Protons with $\SI{1.4}{\giga\electronvolt}$ produce radioisotopes through nuclear reactions in the target material, where tin isotopes are vaporized and then laser-ionized through resonance laser ionization (laser scheme in insert a). Here, IP stands for the ionization potential of tin, and AI denotes the energy for auto-ionizing states. Tin isotopes (red circles) and parasitically ionized contamination (blue and white circles) are extracted at $\SI{30}{\kilo\electronvolt}$ and subsequently mass separated using two electromagnets (High-Resolution Separator, HRS). Ions with the selected mass-over-charge ratio $m/q$ are collected in an ion cooler and buncher (RFQ-cb). The beam is then released in bunches, reduced in energy through the pulsed-drift tube to $\SI{3.2}{\kilo\electronvolt}$, and captured in a time-of-flight spectrometer (MR-ToF MS) at $\SI{2}{\kilo\electronvolt}$. After about $\SI{50}{\milli\second}$, the ions are released, and their flight time to an ion detector is measured. An example ToF spectrum for $m/q=106$ showing strontium-fluoride, indium, and tin is depicted in insert b).}
	\label{fig:apparatus}
\end{figure*}

High-precision Penning-trap mass measurements on neutron-deficient tin isotopes were performed at IGISOL and GSI, reaching down to $^{104}$Sn~\cite{Elomaa_2009_tin, Martin_2007_Sn_measurements}.
More recently, the CSRe storage ring in Lanzhou, China, allowed the direct measurement of $^{103}$Sn for the first time~\cite{2023_Xing_CSRe_measurements} followed by a Penning-trap measurement at FRIB, USA~\cite{ireland2024highprecisionmassmeasurement103sn}, and the TITAN ToF mass spectrometer at {TRIUMF}, Canada, measured down to $^{104}$Sn~\cite{2024_Czihaly_TITAN_measurements}.

Investigations of local trends of the mass surface surrounding $^{100}$Sn have shown several inconsistencies when extracting masses from $Q_{\textrm{EC}}$-value measurements combined with masses of their daughter nuclei~\cite{2021_Mougeot}.
The mass for $^{100}$Sn itself is known only through two inconsistent $Q_{\textrm{EC}}$-value measurements~\cite{Hinke2012_fullauthors, 2019_Lubos_100Sn_gamowteller_fullaothors} in combination with the mass of $^{100}$In~\cite{2021_Mougeot}. 
Furthermore, the mass of $^{103}$Sn, similarly extracted through a $Q_{\textrm{EC}}$-value, was not included in the latest Atomic Mass Evaluation (\texttt{AME2020}~\cite{AME2020_Input_Evaluation}) due to its conflicting impact on the mass surface and a value derived from systematic trends was recommended instead.

In the present work, we report precision mass measurements of $^{103-106}$Sn performed by multi-reflection time-of-flight mass spectrometry at CERN-ISOLDE.
We compare the results with nuclear \textit{ab initio}~\cite{Stroberg:2016ung, Stroberg2019} and density functional theory calculations~\cite{EDF_book} and suggest several adjustments of neighboring masses based on this comparison and local binding energy trends in the region.

\section{\label{sec:Methods}Experiment and Data Analysis}
\subsection{\label{sec:production}Radioactive tin production at ISOLDE}

Neutron-deficient radioactive tin isotopes were produced at CERN-ISOLDE~\cite{2017_Catherall_ISOLDE} through proton-induced spallation of $^{139}$La nuclei,
\begin{align}
    ^{139}\text{La}(p,8p(19+x)n)^{113-x}\text{Sn},
\end{align}
driven by a $\SI{1.4}{\giga\electronvolt}$ proton beam delivered by the CERN Proton Synchrotron Booster (PSB) accelerator.
Figure~\ref{fig:apparatus} shows a schematic of the apparatus for the production, separation, and mass measurement of the radioactive isotopes.  

The target material in the form of lanthanum carbide with a density of $\SI{5}{\gram\per\centi\meter\squared}$ (ISOLDE target unit number \#774) was bombarded by 1 to 2 $\si{\micro\ampere}$ of protons on average.
Each PSB cycle, with a maximum repetition rate of one pulse every $\SI{1.2}{\second}$, contained between 1-3$\times 10^{13}$ protons per cycle.  
Reaction products diffused through and effused out of the porous target material into a small tantalum tube, connecting the target container to the hot-cavity resonance ionization laser ion source (RILIS)~\cite{2017_Fedosseev_RILIS}.
A three-step resonance ionization scheme was used to selectively ionize tin isotopes, shown in insert a) in the center of Fig.~\ref{fig:apparatus}.

Different low-lying electronic states in the tin atom were populated as the radioactive isotopes were produced in the target container heated up to $\SI{1900}{\celsius}$.
To maximize ionization efficiency, two first excitation steps were used in parallel, starting from the \mbox{$5s^2 5p^2$ $^3P_0$} ground state and the first thermally populated state \mbox{$5s^2 5p^2$ $^3P_1$}.
Both transitions led to the \mbox{$5s^2 5p 6s$ $^3P_1$} state at $\SI{34914.28}{\per\centi\meter}$, and the light used to excite them was generated via frequency-tripling a titanium:sapphire (Ti:Sa) laser to produce $\SI{286.4}{\nano\meter}$ and frequency-doubling a dye laser to produce $\SI{301.0}{\nano\meter}$.
The thermal population of the two levels strongly depends on the temperature of the ion source but is almost equal at $\SI{1900}{\celsius}$.
During the experiment, an ionization-efficiency enhancement of roughly 70\% could be observed when using the transition from the thermally populated state in addition to the excitation of the ground state.
A fundamental Ti:Sa laser drove the second step at $\SI{811.6}{\nano\meter}$, exciting the \mbox{$5s^{2}5p6s$ $^3P^o_1 \rightarrow 5s^{2}5p6p$ $^3P_2$} transition.  
An autoionizing transition at $\SI{823.7}{\nano\meter}$, driven by a Ti:Sa laser, was used as the final step.

The ionized tin isotopes were then extracted from the target and ion source front-end and accelerated as a beam of singly-charged radioactive ions at $\SI{30}{\kilo\electronvolt}$.  
Tin ions of interest were separated from the extracted ion beam based on their mass-over-charge ratio $m/q$ using the High Resolution Separator (HRS), which consists of two successive electromagnets depicted in dark grey in Fig.~\ref{fig:apparatus}.
The tin ions were finally delivered to the ISOLTRAP mass spectrometer~\cite{2008_Mukherjee}, shown at the bottom of Fig.~\ref{fig:apparatus}.

\subsection{\label{sec:mass_meas}Mass measurement method}

At the mass spectrometer, the ion beam was firstly decelerated and captured in a linear radio-frequency quadrupole cooler/buncher (RFQ-cb)~\cite{Herfurth2001}, using helium buffer-gas cooling to improve beam emittance and to create a bunched beam with temporal bunch sizes in the order of several tens of nanoseconds. 
After a cooling time of $\SI{10}{\milli\second}$, the ion bunch was released from the trap and re-accelerated to $\SI{30}{\kilo\electronvolt}$.
To capture the ion bunch in the multi-reflection time-of-flight mass spectrometer (MR-ToF MS)~\cite{2013_wolf}, the ion kinetic energy was first reduced from $\SI{30}{\kilo\electronvolt}$ to $\SI{3.2}{\kilo\electronvolt}$ using a pulsed drift tube between the RFQ-cb and the MR-ToF MS, and once more from $\SI{3.2}{\kilo\electronvolt}$ to $\SI{2}{\kilo\electronvolt}$ using a pulsed drift tube in-between the two electrostatic mirrors of the mass spectrometer.
Trapped inside the device, the ion bunch was reflected between the mirrors for up to 2000 turns before being ejected using the same pulsed drift tube, the so-called in-trap lift~\cite{2017_Wienholtz}.
The ion ToF was then measured using a {MagneTOF\textsuperscript{\texttrademark} Mini 14924} electron multiplier detector downstream of the device.

The atomic mass $m$ of the ions can be extracted using the measured time of flight $t$ with respect to two reference masses $m_1$ and $m_2$ and their ToFs $t_1$ and $t_2$ as:
\begin{align}
\sqrt{m}=C_{\text{ToF}}\Delta_{\text{Ref}}+\Sigma_{\text{Ref}}/2, 
\label{eq:CTOF}
\end{align}
where $\Delta_{\text{Ref}}=\sqrt{m_1}-\sqrt{m_2}$, $\Sigma_{\text{Ref}}=\sqrt{m_1}+\sqrt{m_2}$ and $C_{\text{ToF}}=(2t-t_1-t_2)/\left[2(t_1-t_2)\right]$~\cite{2013_Wienholtz}.
As reference ions, the isobaric indium isotope from the same spectrum and $^{87}$Rb$^+$ from interleaved reference measurements provided by a surface-ionization source at ISOLTRAP have been used.
The mass-resolving power of the device is defined as
\begin{align}
R=\frac{m}{\Delta m} = \frac{t}{2\Delta t_\text{FWHM}}, \label{eq:mass_resolving_power_MRTOFMS}
\end{align}
where two different components in the ToF spectrum are considered fully resolved when their separation is at least twice the full-width-half-maximum~(FWHM) of their ToF distribution.

The MR-ToF MS was operated in near-isochronous mode, in which faster ions penetrate the reflecting mirror potential deeper than slower ions, leading to longer turn-around times, thus compensating for their higher velocities.
In such a way, the time focus was retained for long storage times, i.e. up to $\SI{50}{\milli\second}$, hence improving resolving power and resolution.
This mode of operation was achieved by precisely tuning the ion kinetic energy inside the device by means of the in-trap lift voltage during pulsing.
An example ToF spectrum for $m/q=106$ is depicted in insert b) of Figure~\ref{fig:apparatus}.

% \begin{figure*}[tb!]
%   \centering
%   \subfloat[Normalized yields of the extracted $^{106}$Sn$^+$ beam as measured throughout the experiment with respect to target and ion source temperatures. Shaded areas indicate where protons from the PSB were not available. The grey dashed line indicates the sublimation temperature of the target material.]{\includegraphics[width=0.58\linewidth]{yield_vs_time.pdf}}
%   \hfill
%   \subfloat[Normalized yield and beam purity of the extracted $^{106}$Sn$^+$ beam versus ion source temperature. The target temperature in the shaded area was $\SI{1650}{\degree\celsius}$, in the nonshaded area $\SI{1730}{\degree\celsius}$.]
%   {\includegraphics[width=0.42\linewidth]{yield_vs_line.pdf}}
%   \caption[Target behavior]{Behavior of the yield of $^{106}$Sn versus target and line temperature, as well as over time. Decreasing yields, most likely caused by sintering of the target material, can be observed towards the end of the experiment.}
%   \label{fig:target_performance}
% \end{figure*}

\subsection{\label{sec:target}Target and ion source performance}

\begin{figure*}[t!]
    \includegraphics[width=\textwidth]{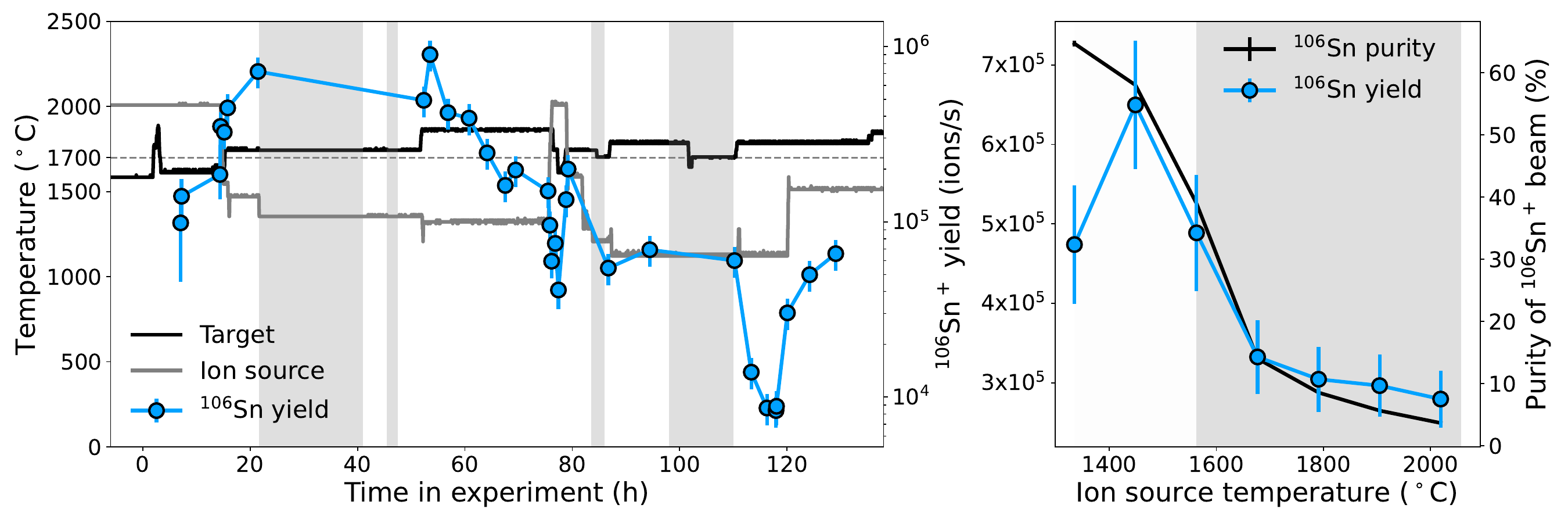}
    \caption{Left: Normalized yields (in ions per second normalized to $\SI{1}{\micro\ampere}$ of protons on target) of the extracted $^{106}$Sn$^+$ beam as measured by ISOLTRAP throughout the experiment with respect to target and ion source temperatures.
    The temperatures were calibrated using an optical pyrometer for a range of applied heating currents and powers, with measurement uncertainties of about $\SI{50}{\celsius}$ (not shown).
    Shaded areas indicate where protons from the PSB were not available.
    The grey dashed line indicates the temperature at which the target material starts to undergo sintering.
    Right: Normalized yield and beam purity (percentage of tin ions in the MR-ToF spectrum) of the extracted $^{106}$Sn$^+$ beam versus ion-source temperature.
    The target temperature in the shaded area was $\SI{1650\pm 50}{\celsius}$, in the nonshaded area $\SI{1730\pm 50}{\celsius}$.
    \label{fig:target_performance}}
\end{figure*}

Using the MR-ToF MS, the yield of the extracted neutron-deficient tin isotopes has been tracked over the period of the experiment using $^{106}$Sn as a reference to monitor possible target micro-structure degradation and stability of the RILIS.
Furthermore, $^{106}$Sn$^+$ served as a test case to optimize target and ion source heating parameters with respect to extraction efficiency and beam purity.
Figure~\ref{fig:target_performance} (left) shows the normalized yield of the extracted $^{106}$Sn$^+$ versus time, together with target and ion source heating values. 
Yields in the order of $10^5$ to $10^6$ ions/s could be achieved throughout most of the experiment.
However, towards the end, yields dropped on average by almost one order of magnitude.
% The grey-shaded areas in Fig.~\ref{fig:target_performance} (left) indicate when protons delivered by the PSB were not available.

At the beginning of the experiment, target and ion source heating were optimized based on the $^{106}$Sn$^+$ signal, shown in Fig.~\ref{fig:target_performance} (right). 
Lower ion source temperatures strongly suppressed isobaric contamination, such as indium and strontium monofluoride, which are parasitically surface-ionized in the hot environment of the ion source and transfer line.
The highest yields and lowest contamination levels were observed for increased target heating and lower ion source temperatures. 
High target temperatures facilitate the diffusion and effusion of reaction products through the target material and into the ion source. 
In contrast, a lower ion source temperature reduces the degree of parasitic surface ionization, improving the beam purity.

Besides the decrease in positive ion density, the thermal emission of electrons from the walls of the ion source decreases with lower temperature, while a surplus of negative charges is detrimental for ion confinement~\cite{1981_kirchner_ion_sources, 2012_Turek_hot_ion_source}.
% Also, the transversal ion confinement by a plasma sheath and, thus, ion survival probability in the ion source is affected.
Additionally, a lower heating current implies a smaller ohmic voltage gradient along the ion source, which leads to a slower drift of ions towards the extraction electrode.
% The systematic investigations show that, in this case, the surplus of surface ionization with higher temperature dominates these effects.
%This reduces the amount of space charge in the ion source, i.e. the amount of ions from the target and the number of electrons emitted from the walls of the ion source container, which generally create the confining potential. 
% A lower heating current also results in a smaller potential across the ion source, which leads to a slower drift of ions towards the extraction electrode.
The increase in yield at lower ion source temperatures in Fig.~\ref{fig:target_performance} (right) is most likely due to a slightly lower target temperature ($\SI{1650\pm 50}{\celsius}$) as compared to the higher temperatures in the shaded areas of the figure ($\SI{1730\pm 50}{\celsius}$).

Due to the short half-lives of the tin isotopes of interest, the target was operated at or above $\SI{1700\pm 50}{\celsius}$ to promote a quick diffusion and effusion from the target material.
However, a drop in yield over time was observed, which can be attributed to the operation of the target container at temperatures at which lanthanum carbide undergoes a substantial degree of sintering, which leads to a decrease in the target material's open porosity~\cite{2008_Biasetto} and results in a reduction of the effusion rate of the radioactive isotopes.
Furthermore, the sublimation temperature of pure LaC$_{2}$ in vacuum is near $\SI{1700}{\celsius}$, at which the target material starts to volatilize, further worsening the target's release properties.
Additionally, the energy deposited by the pulsed proton beam leads to mechanical stress and a local hotspot, accelerating the deterioration of the target microstructure.

\begin{figure}[t]
    \centering
    \includegraphics[width=0.5\columnwidth]{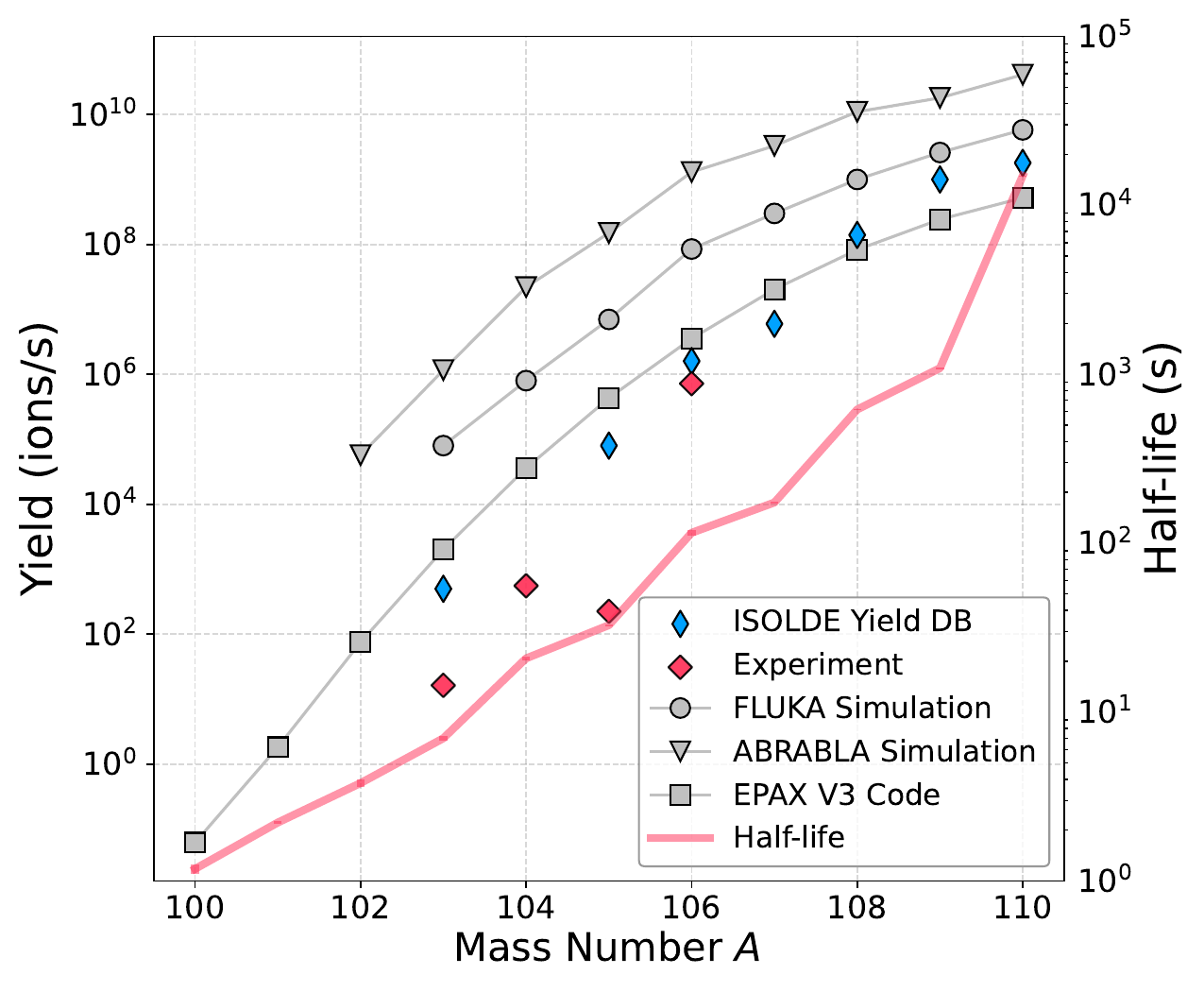}
    \caption{Best yields extracted from the target measured with the MR-ToF MS (red) and normalized to a proton current of $\SI{1}{\micro\ampere}$. In comparison, previously measured yields of extracted beams reported in the ISOLDE yield database (blue) and in-target production simulation codes FLUKA~\cite{2014_Bohlen_FLUKA, Ahdida_2022_FLUKA} and ABRABLA~\cite{2009_Kelic_ABRABLA}, as well as the empirical fragmentation production cross section code EPAX-V3~\cite{2012_Summerer_EPAX_V3}.
    \label{fig:yields}}
\end{figure}

Figure~\ref{fig:yields} shows the best yields for observed neutron-deficient tin isotopes.
These are compared with previously reported values from the ISOLDE yield database~\cite{yield_db, 2020_Ballof_ISOLDE_Yield_DB}, to in-target production simulations (available in the same database)~\cite{2014_Bohlen_FLUKA, Ahdida_2022_FLUKA, 2009_Kelic_ABRABLA}, and to empirical in-target production cross-section calculations~\cite{2012_Summerer_EPAX_V3}.
On average, the yields were significantly lower than previously reported best yields from the same target and ion source combination at ISOLDE (compare red and blue diamonds)~\cite{yield_db, 2022_CRIS_in_prep}.
The measured yield for $^{105}$Sn$^+$ is disproportionately lower than the trend would suggest, most likely caused by a low ion source temperature of $\SI{1150\pm 50}{\celsius}$ used during that specific measurement, which was also observed to have a negative impact on the extracted yield of $^{106}$Sn (see Fig.~\ref{fig:target_performance} (right)).

The trends and values of the different production simulations for the tin isotopes closer to stability match the measurements within one order of magnitude on average, as most of the produced isotopes in the target can be extracted and ionized due to their long half-lives.
For isotopes with half-lives well below $\SI{1e4}{\second}$, the trend of the extracted ions deviates significantly from the calculated in-target production cross-section, indicating large losses during the extraction process due to target-release characteristics, ionization efficiency, and radioactive decay.
These losses become significant below half-lives of $\SI{100}{\second}$, where the extracted yields differ by more than two orders of magnitude, highlighting the challenges of diffusion of isotopes inside the target material in the ISOL production method~\cite{2001_Koester_ISOL}.

\subsection{\label{sec:spectra}ToF spectra and fitting}

\begin{figure}[t]
    \includegraphics[width=0.5\columnwidth]{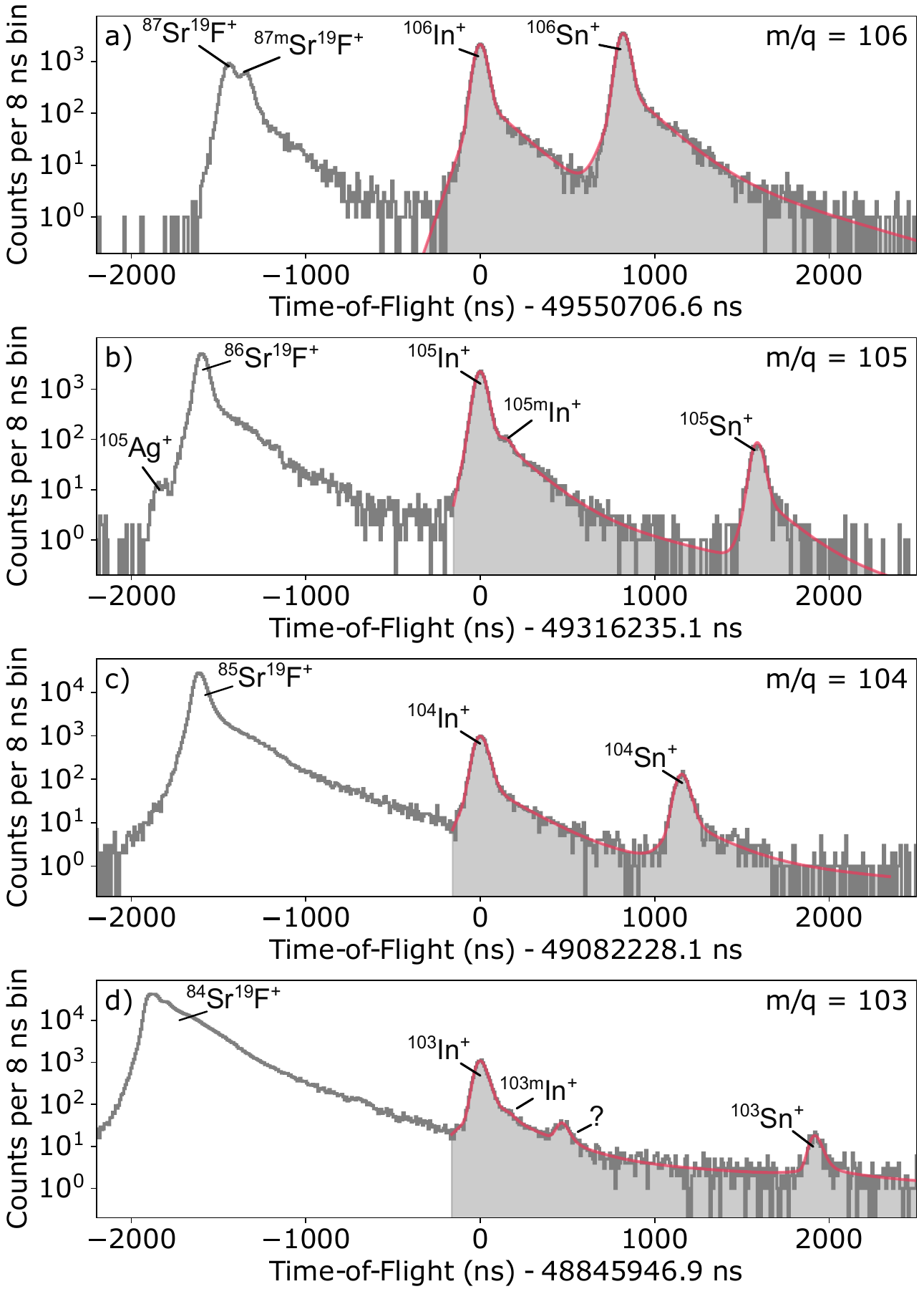}
    \caption{Typical ToF spectra for the four tin isotopes investigated in the present study. The grey-shaded area indicates the fit range, which excludes the potentially saturated peaks at smaller ToFs. The red line shows the fit to the data. The horizontal axis is centered at the indium peak. 
    \label{fig:tof_spectra}}
\end{figure}

Figure~\ref{fig:tof_spectra} shows ToF data of the four different mass measurements presented in this study. 
In all spectra, the surface-ionized strontium-fluoride and indium contaminants are present. 
While for $m/q=106$, this contamination could be reduced to less than $40\%$ (shown in Fig.~\ref{fig:target_performance} (right)), for smaller $m/q$ values, the contamination gradually increased to $>$99.9$\%$ of the isobaric beam.

To calculate the atomic masses as shown in Eq.~\eqref{eq:CTOF}, the Tof was extracted by fitting a multi-component exponentially modified Gaussian distribution (``hyper-EMG'')~\cite{2017_Purushothaman_hyperEMG} to the data.
Fits with different components were tested to determine the necessary number of components for the fit function, and the resulting reduced chi-squared $\chi^2_\text{red}$ was observed.
One negative component and three positive components resulted in $\chi^2_\text{red}<10$, sufficiently representing the data without leading to parameter overfitting.
The final fit was performed using unbinned maximum likelihood estimation, implemented in CERN's RooFit MINUIT minimizer~\cite{James:2296388}.   
With 2000 revolutions resulting in storage times of up to $\SI{50}{\milli\second}$ in the MR-ToF MS, mass resolving powers $R$ of up to $300\,000$ were achieved during the experimental campaign, which proved sufficient to resolve the isomeric states in the odd-even indium isotopes and in $^{87}$Sr (with excitation energies of about $\SI{650}{\kilo\electronvolt}$ and $\SI{390}{\kilo\electronvolt}$, respectively).

Due to the intense contamination levels on $m/q=104$ and $m/q=103$, only the ToF range between indium and tin was fitted for all masses.
This prevented systematic effects resulting from detector and data acquisition saturation around the arrival time of the SrF$^+$ ions from being introduced into the shape of the fit function.
For the fitted time range, the average count rate was well below five ions per experimental cycle, which prevented detector pileup (the single-ion detector response of the detector is about $\SI{1.5}{\nano\second}$ long for a MagneTOF\textsuperscript{\texttrademark} Mini 14924) nor to loss of signals due to data acquisition deadtime ($<$$\SI{100}{\pico\second}$ for FAST ComTec MCS6A).
Furthermore, the ion load in the MR-ToF MS was kept well below 20 ions per cycle to avoid space-charge effects that can lead to systematic ToF shifts~\cite{Maier2023spacecharge}.

In Fig.~\ref{fig:tof_spectra}a, an example of the spectra for $m/q=106$ is shown, where one can see, from left to right, the contamination $^{87}\text{Sr}^{19}\text{F}^+$ and $^{106}\text{In}^+$, as well as the isotope of interest $^{106}\text{Sn}^+$.
Interestingly, as noticed during the systematic measurements of this mass-over-charge ratio throughout the experiment, the shoulder on the $^{87}\text{Sr}^{19}\text{F}^+$ peak only appeared after several tens of hours into the experiment.
The observed ToF matches that of a strontium-fluoride molecule containing the long-lived $I=1/2$ isomeric state with an excitation energy of $\SI{389}{\kilo\electronvolt}$, leading to a slightly longer ToF.
The involved time scales of the isomer's appearance suggest that the strontium isomer is not directly produced through nuclear reactions but rather through the in-target beta-decay of the $I=1/2$ ground state of $^{87}$Y with a half-life of almost $T=\SI{80}{\hour}$~\cite{87Sr_decay_NDS}. 
A typical measurement for the beam composition on this $m/q$ value took five minutes.
Figure~\ref{fig:tof_spectra}a shows data for about $\SI{20}{\minute}$ of measurement time.

Figure~\ref{fig:tof_spectra}b shows the data taken for $m/q=105$, where $^{105}\text{Ag}^+$, $^{86}\text{Sr}^{19}\text{F}^+$, $^{105g,m}\text{In}^+$, and $^{105}\text{Sn}^+$ were present in the beam. 
In this spectrum, the $I=1/2$ isomer of $^{105}\text{In}$ was observed with an abundance of about 1:80 with respect to the $I=9/2$ ground state, and the spectrum was fitted with three individual peaks.   
The data for this spectrum corresponds to a measurement time of about $\SI{800}{\second}$.

For the ion beam with $m/q=104$, $^{85}\text{Sr}^{19}\text{F}^+$, $^{104}\text{In}^+$, and $^{104}\text{Sn}^+$ were observed, and the spectrum was fitted with only two individual peaks. 
Interestingly, the ${I=1/2}$ isomeric state in the $^{85}\text{Sr}$ containing molecule was not observed, despite the predominant beta-decay of $^{85}\text{Y}$ into the isomeric state~\cite{85Sr_decay_NDS}.
% This might be because the excitation energy of the isomer with about $\SI{239}{\kilo\electronvolt}$ was too small to be resolved by the spectrometer.
The data within this plot was taken in about $\SI{1}{\hour}$ of measurement.

\begin{table*}[t!]
\caption[Mass results]{Mass measurement results in comparison with storage ring measurements from~\cite{2023_Xing_CSRe_measurements}, MR-ToF MS measurements from~\cite{2024_Czihaly_TITAN_measurements}, and the \texttt{AME2020}, which is based on the Penning trap measurements reported in~\cite{Elomaa_2009_tin}. Spin assignments $J^\pi$, half-lives, and reference masses ($^A$In for mass $m_1$, and $^{87}$Rb for mass $m_2$) are taken from the \texttt{AME2020}. Values marked with $\#$ are extrapolated or assigned from systematics.}	
    {\color{black}
	\begin{tabular}{ l c c c c c c c c c} \hline\hline 
	    & & & & & \multicolumn{4}{c}{Mass excess ($\si{\kilo\electronvolt}$)} \\
		A & J$^\pi$ & Half-life & Ref. ions & C$_{\text{ToF}}$ & This work & \texttt{AME2020}~\cite{AME2020} & Lanzhou~\cite{2023_Xing_CSRe_measurements} & TITAN~\cite{2024_Czihaly_TITAN_measurements} \\ \hline 
		103 & $5/2^{+}$\# & $\SI{7.0\pm0.2}{\second}$ & \ce{^{87}Rb+}, \ce{^{103}In+} & $\SI{0.500484342\pm 0.000000920}{}$ & $\SI{-67104\pm 18}{}$ & $\SI{-67090\pm 100}{}\#$ & $\SI{-67138\pm 68}{}$ & / \\
		104 & $0^{+}$\# & $\SI{20.8\pm0.5}{\second}$ & \ce{^{87}Rb+}, \ce{^{104}In+} & $\SI{0.500275256\pm 0.000000476}{}$ & $\SI{-71629\pm 10}{}$ & $\SI{-71627\pm6}{}$ & / & $\SI{-71601\pm 50}{}$\\
		105 & $5/2^{+}$\# & $\SI{32.7\pm0.5}{\second}$ & \ce{^{87}Rb+}, \ce{^{105}In+} & $\SI{0.500358319\pm0.000000329}{}$ & $\SI{-73348\pm 13}{}$ & $\SI{-73338\pm 4}{}$ & / & $\SI{-73349\pm 34}{}$\\
		106 & $0^{+}$\# & $\SI{1.92\pm0.08}{\minute}$ & \ce{^{87}Rb+}, \ce{^{106}In+} & $\SI{0.500175627\pm0.000000071}{}$ & $\SI{-77346\pm 14}{}$ & $\SI{-77354\pm 5}{}$ & / & $\SI{-77327\pm 37}{}$ \\
		\hline
	\end{tabular}}
	\label{tab:results}
\end{table*}

On $m/q=103$, one can notice the increase of contamination in the beam: the $^{84}\text{Sr}^{19}\text{F}^+$ peak on the left is saturating the discriminator-based data acquisition (note the double-peak structure and the pronounced tailing). 
Again, both isomer and ground state in $^{103}\text{In}^+$ were resolved.
However, one additional peak appears between the two indium states and $^{103}\text{Sn}^+$ on the right-hand side of the spectrum, which is inconsistent with any isobaric mass or previously observed surface-ionized molecule.
It can not be ruled out that this species was ejected from the MR-ToF MS on a different revolution number than the other isotopes, hence not having traveled the same distance.
Such contamination can usually be investigated by varying the overall trapping time and observing the relative number of peaks in the spectra. 
This was not attempted due to the limited time available for these $m/q$ settings.
As mentioned in Section~\ref{sec:target}, due to the high operation temperature and proton current during long periods of time, which most likely led to target material sintering, yields for the more short-lived tin isotopes dropped quickly throughout the experiment.
As a result, the ratio of $^{103}\text{Sn}^+$ versus contamination in the beam worsened throughout the experiment, preventing the measurement of the lighter neutron-deficient tin isotopes. 
Effectively, $\SI{2.4}{\hour}$ of data from the first day after the start of the experiment are included in the plot and the analysis of the final mass value.

\section{\label{sec:results_discussion}Results and discussion}

Table~\ref{tab:results} shows the results for $^{103-106}$Sn, together with available literature values on mass excesses and half-lives, as well as tentative spin assignments.
The references for extracting the mass of the ion of interest were consistently taken from the isobaric indium ions and from $^{87}$Rb$^+$ provided by the ISOLTRAP offline ion source. 
All mass excess values are in very good agreement with the available literature data of the \texttt{AME2020}~\cite{AME2020}, the storage ring measurements from Lanzhou~\cite{2023_Xing_CSRe_measurements}, and the MR-ToF MS measurements from TITAN at TRIUMF~\cite{2024_Czihaly_TITAN_measurements}.
The experimental uncertainty is between $\SI{10}{\kilo\electronvolt}$ and $\SI{20}{\kilo\electronvolt}$, reaching a precision of about ${\delta m/m \sim 2\times10^{-7}}$.
The reported mass on $^{103}$Sn is a factor four more precise than the storage ring measurement ($\SI{-67138\pm 68}{\kilo\electronvolt}$)~\cite{2023_Xing_CSRe_measurements} and, in combination with the other mass measurements, allows evaluation of the mass surface in the neutron-deficient tin region.
With this new measurement, we can deduce the $Q_{\textrm{EC}}$ value for the decay of $^{103}$Sn into $^{103}$In to be $\SI{7528\pm 20}{\kilo\electronvolt}$, compared to $\SI{7640\pm 70}{\kilo\electronvolt}$ reported in~\cite{Kavatsyuk2005} which was extracted from total absorption spectroscopy and beta-delayed proton energy measurements.

To investigate the impact of the shift and increased precision of the $^{103}$Sn mass measurement on the local mass surface, the two-neutron separation energy
\begin{align}
    S_{2n}(Z,N) = B(Z, N - 2) - B(Z, N)
\end{align}
and the odd-even staggering estimator
\begin{align}
    \Delta_{3n}&=0.5\cdot(-1)^N\cdot [B(Z,N-1) \notag \\
    &-2\cdot B(Z,N)+ B(Z,N+1)],
\end{align}
are calculated, where $B(Z,N)=M_E(N, Z) - Z \cdot M_E(^1\text{H}) - N \cdot M_E(n)$ is the total nuclear binding energy which consists of the nuclear mass excess $M_E(N,Z)$ of the nucleus of interest, and the mass excesses of the hydrogen atom $^1\text{H}$ and the mass excess of the neutron.
The $S_{2n}$ filter is regularly used to evaluate shell closures and the sudden change of nuclear deformation.
Typically, one analyzes the change of slope in $S_{2n}$ values with the addition or removal of neutrons, especially at magic numbers.
%and deformation-dependent changes in binding energies, highlighted by sudden changes in the slope between neighboring neutron numbers, especially at magic numbers.
Complementarily, the $\Delta_{3n}$ filter quantifies the strength of the odd-even staggering along an isotopic chain.

\begin{figure}[t!]
    \includegraphics[width=0.5\columnwidth]{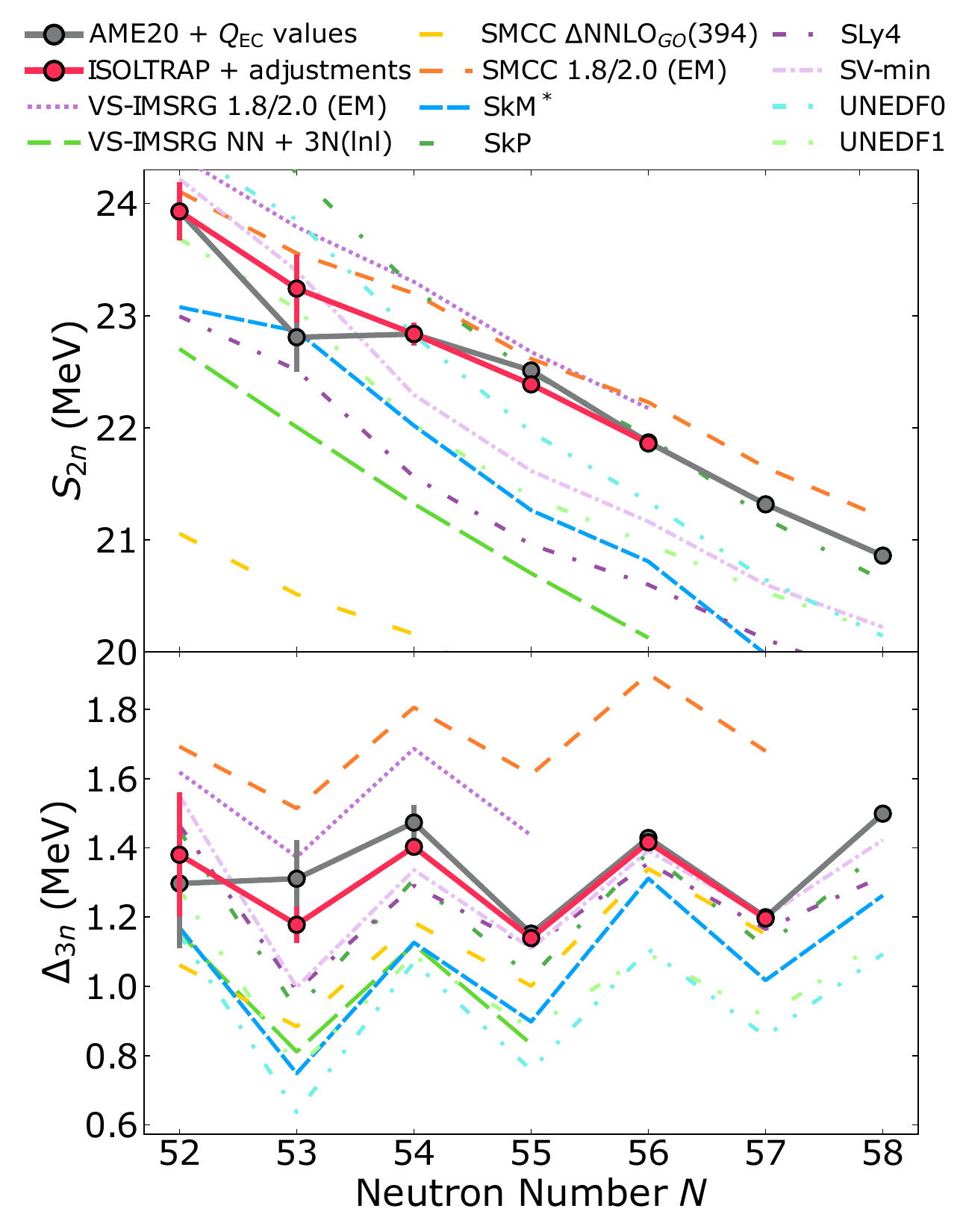}
    \caption{Evaluated two-neutron separation energy (top) and odd-even staggering estimator (bottom) from masses reported in the \texttt{AME2020} (grey), the ISOLTRAP measurement with adjustments (red), nuclear \textit{ab initio} calculations from~\cite{2021_Mougeot}, and DFT calculations from~\cite{Erler2012}. The grey data points include the masses of $^{101-103}$Sn calculated with the $Q_{\textrm{EC}}$ value measurements from~\cite{Straub2010, Faestermann2002, Kavatsyuk2005}. The red points include this work's mass measurement of $^{103}$Sn and the extrapolated mass value for $^{101}$Sn.
    \label{fig:S2n_theory}}
\end{figure}

The two mass filters are plotted in Fig.~\ref{fig:S2n_theory}, comparing mass values extracted based on $Q_{\textrm{EC}}$ values for $^{101-103}$Sn (grey) and direct mass measurements for $^{103-108}$Sn (red) with nuclear \textit{ab initio} calculations from~\cite{2021_Mougeot} and density functional theory (DFT) calculations described in Ref.~\cite{Erler2012} and references therein.
In the upper panel, using the $S_{2n}$ filter, one can observe a smooth evolution in the binding energy differences above the expected shell closure at ${N=50}$.
This smooth trend is irregular at ${N=53}$ and at ${N=55}$ when using masses for $^{101-103}$Sn derived using $Q_{\textrm{EC}}$ links.
This irregularity is further highlighted by plotting the odd-even staggering estimator, shown in the lower panel of Fig.~\ref{fig:S2n_theory}.
% The staggering between ${N=52}$ and ${N=54}$ is recovered through the adjustment, as predicted by the theoretical calculations.

Extracting masses through $\beta$ end-point energies requires complete knowledge of the feeding into excited states in the daughter nucleus to reliably fit the $\beta$-decay spectrum.
High-resolution $\gamma$-spectroscopy of the decay is needed to account for all internal transitions and total absorption spectroscopy, as well as energy measurements of the delayed proton emission to draw a full picture of the decay.
However, incomplete knowledge of the full decay chain caused by internal $\gamma$ conversion or undetected weak decay branches may lead to significant systematic errors.
In the 2020 Atomic Mass Evaluation of input data~\cite{AME2020_Input_Evaluation}, the mass of $^{101}$Sn is extracted through a $\beta$-delayed proton decay channel via $^{101}$In into $^{100}$Cd, with a dominating mass uncertainty of $\SI{300}{\kilo\electronvolt}$ assigned from a measurement of the proton emission energy and branching ratios, reported in a scientific report at GSI~\cite{Straub2010}.
Similarily, the ground-state mass of $^{102}$Sn is extracted from the $^{102}$In ground-state mass using the $Q_{\textrm{EC}}$ value measured in~\cite{Faestermann2002}, incorporating the $\SI{100}{\kilo\electronvolt}$ measurement uncertainty.
Additionally, $^{103}$Sn was deemed a "seriously irregular mass" determined only from the $^{103}$In ground-state mass through the $Q_{\textrm{EC}}$ value measurement from Ref.~\cite{Kavatsyuk2005}, which led to its rejection from the \texttt{AME2020}~\cite{AME2020_Input_Evaluation}.
Finally, the precise mass measurement of $^{100}$In from Ref.~\cite{2021_Mougeot} highlighted another discrepancy, namely the competition between the results of the two direct beta-decay measurements of $^{100}$Sn~\cite{Hinke2012_fullauthors, 2019_Lubos_100Sn_gamowteller_fullaothors}, whose $Q_{\textrm{EC}}$ values disagree on a $2\sigma$ uncertainty level.  

Based on the present precise mass measurements, we reevaluate the mass values expected from systematic trends and from comparing them to theoretical calculations.
The present $^{103}$Sn mass excess of $\SI{-67104\pm 18}{\kilo\electronvolt}$ makes this nucleus about $\SI{130}{\kilo\electronvolt}$ less bound compared to the value of $\SI{-66970\pm 70}{\kilo\electronvolt}$ extracted from the $Q_{\textrm{EC}}$ value.
However, to fully remove the kink at ${N=53}$, one must also adjust the mass of $^{101}$Sn, which contributes in equal parts to the $S_{2n}$ value.
By adding $\SI{+300}{\kilo\electronvolt}$ to the mass excess of $^{101}$Sn, within two standard deviations of the reported beta-delayed proton energy measurement~\cite{Straub2010}, one obtains a smooth trend, as highlighted by the red data point in Fig.~\ref{fig:S2n_theory}.
We motivate this adjustment not only by comparing the local trends of the mass surface near the shell closure in ${N=50}$ but also based on the predicted trends by the nuclear \textit{ab initio} and DFT calculations. 
While not all methods, interactions, and functionals predict the correct magnitude of the $S_{2n}$ value, all calculations predict a flat trend.

\begin{table}[b!]
\caption[Mass results]{Mass excess values derived from the extrapolated mass of $^{101}$Sn and present measurements of $^{103}$Sn combined with Ref.~\cite{2023_Xing_CSRe_measurements}. The re-evaluated mass excess values are compared to those reported in the \texttt{AME2020}~\cite{AME2020}. 
Values marked with $\#$ are extrapolated from the mass surface, and values marked with $*$ are based on the present adjustment of $^{101}$Sn.}	
\begin{tabular}{l c c} \hline\hline 
        & Mass excess ($\si{\kilo\electronvolt}$)& \texttt{AME2020} ($\si{\kilo\electronvolt}$)\\ \hline
        $^{101}$Sn & $\SI{-60005\pm 300}{}*$ & $\SI{-60305\pm 300}{}\#$ \\
        $^{103}$Sn & $\SI{-67105\pm 16}{}$ & $\SI{-67092\pm 100}{}\#$ \\
        $^{104}$Sb & $\SI{-59307\pm 22}{}$ & $\SI{-59340\pm 100}{}\#$ \\
        $^{105}$Te & $\SI{-52510\pm 300}{}*$ & $\SI{-52810\pm 300}{}\#$\\
        $^{107}$Te & $\SI{-60670\pm 17}{}$ & $\SI{-60660\pm 100}{}\#$\\
        $^{108}$I & $\SI{-52784\pm 21}{}$ & $\SI{-52770\pm 100}{}\#$ \\
        $^{109}$Xe & $\SI{-45870\pm 300}{}*$ & $\SI{-46170\pm 300}{}\#$\\
        $^{111}$Xe & $\SI{-54530\pm 60}{}$ & $\SI{-54520\pm 120}{}\#$\\
        $^{112}$Cs & $\SI{-46430\pm 60}{}$ & $\SI{-46415\pm 120}{}\#$\\
    \hline
\end{tabular}
\label{tab:AMEupdate}
\end{table}

Contrary to extracting masses through beta-decay endpoint energies, nuclear $\alpha$-decay is an intrinsically accurate method as the kinetic energies of the emitted alpha-particle are mono-energetic and relatively easy to measure with $\si{\kilo\electronvolt}$ precision. 
A unique feature of the nuclear landscape is the small island of $\alpha$-emitters with decay chains ending in neutron-deficient tin isotopes.
Most of the mass data of those short-lived isotopes are indirectly known based on anchoring tin masses through successive $\alpha$-decays by measuring the $\alpha$-particles energies, see for example Ref.~\cite{Darby_2010_alpha_decay_109Xe, Auranen_2018_Alpha_decay_to_100Sn}.
By directly measuring $^{101-103}$Sn, one can significantly reduce the mass uncertainties for antimony, tellurium, iodine, xenon, and cesium isotopes, which are otherwise very challenging for ion-trap-based atomic mass measurements.
Table~\ref{tab:AMEupdate} shows mass excess values, which are based on our $^{103}$Sn mass measurement combined with the measurement of Ref.~\cite{2023_Xing_CSRe_measurements} and the proposed $^{101}$Sn mass adjustment.

\begin{figure}[t]
    \includegraphics[width=0.5\columnwidth]{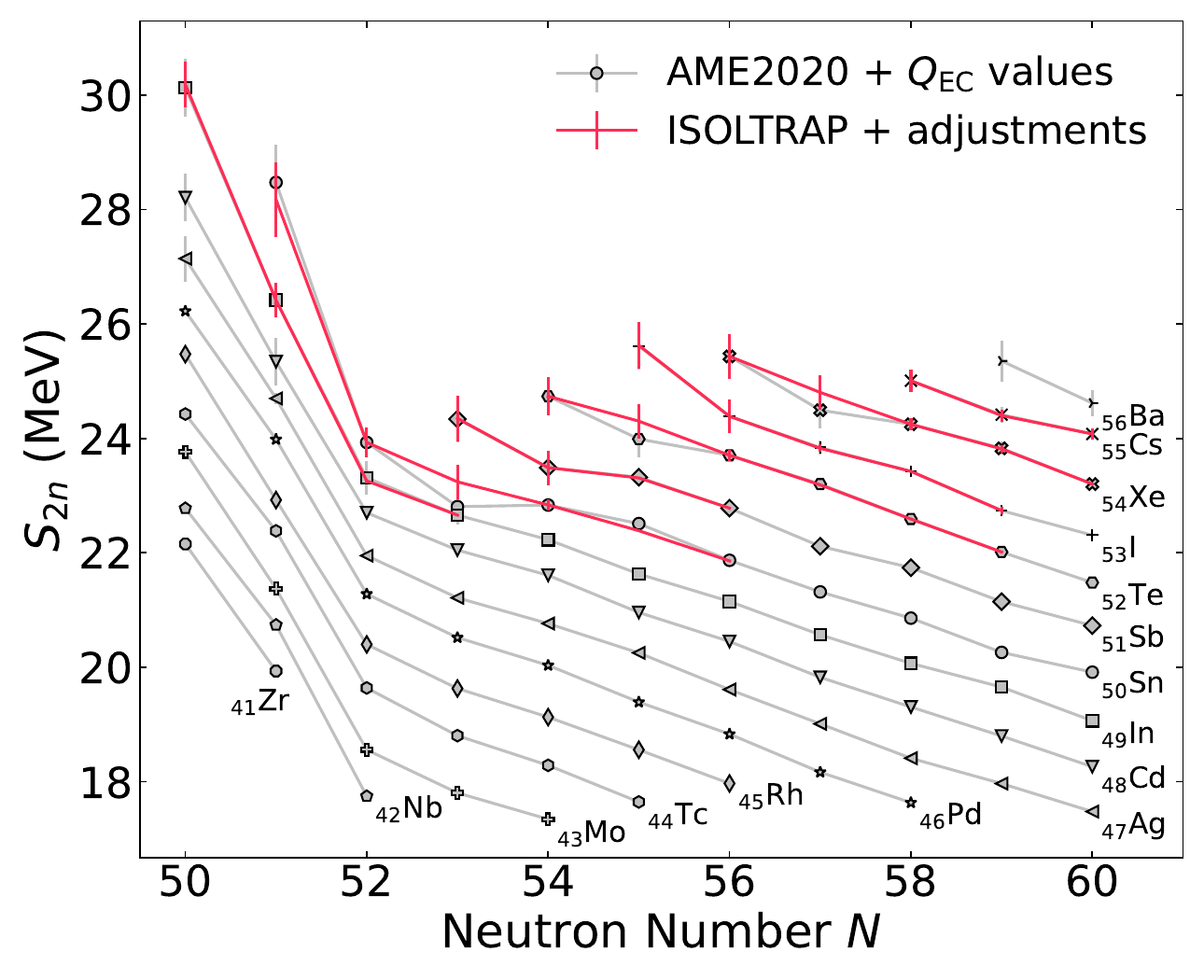}
    \caption{Evaluated two-neutron separation energy from masses reported in the \texttt{AME2020} including the masses of $^{101-103}$Sn calculated from $Q_{\textrm{EC}}$ value measurements~\cite{Straub2010, Faestermann2002, Kavatsyuk2005} (grey) or readjusted as shown in Tab.~\ref{tab:AMEupdate} following the ISOLTRAP mass measurements from this work (red). Plotted improvements on indium mass uncertainties are taken from~\cite{2021_Mougeot, 2023_Nies_In}. 
    \label{fig:S2n}}
\end{figure}

The masses from Tab.~\ref{tab:AMEupdate} in combination with the mass tables from the \texttt{AME2020} can be used to plot the $S_{2n}$ values for a large range of isotopes ($Z=41-56$), shown in Fig.~\ref{fig:S2n}.
On a more global picture, the trends for the $S_{2n}$ values are extremely smooth above the ${N=50}$ shell closure for nuclei below with ${Z<50}$, where all necessary masses for the evaluation are known to sufficient precision. 
This smoothness has now been recovered for the tin chain.
Above ${Z=50}$, one notices significant changes at ${N=55}$ in the tellurium chain and at ${N=57}$ in the xenon chain, leading to overall smoother trends as well. 

\section{\label{sec:Conclusion}Conclusion}

Multi-Reflection ToF Mass Spectrometry was performed on $^{103-106}$Sn isotopes produced at the radioactive ion beam facility at CERN-ISOLDE.
Radioactive tin isotopes were produced through nuclear reactions of high-energy protons impinging on a LaC$_x$ target and were ionized through a three-step laser ionization scheme.
The target and ion source performance was tracked based on yields of $^{106}$Sn$^+$, which dropped with time, likely because of sintering of the target material.
This study offers a detailed basis for future developments of neutron-deficient tin beams at ISOL facilities and strongly motivates further improvements of ISOL target materials and their degradation over time.
% and dropping yields throughout the experiment were observed, likely caused by sintering of the target material.
The highest yields of extracted $^{103-106}$Sn$^{+}$ ions were compared to in-target production simulations and empirical cross-section calculations, showing slightly lower values than previously achieved during other experiments at ISOLDE.

All cases suffered from surface-ionized contamination and target degradation, preventing mass measurements of lighter tin isotopes despite sufficient production rates initially estimated from simulations.
Atomic masses for $^{103-106}$Sn were obtained, with uncertainties ranging between $\SI{10}{\kilo\electronvolt}$ and $\SI{20}{\kilo\electronvolt}$.
The mass uncertainty of $^{103}$Sn was improved by roughly a factor of four.
The measured masses were compared with nuclear \textit{ab initio} and DFT calculations to assess the local smoothness of the mass surface near $^{100}$Sn.
It was shown that the extracted masses for $^{101,103}$Sn from $Q_{\textrm{EC}}$ value measurements are very likely inaccurate and lead to discrepant trends.
Using the more precise mass value of $^{103}$Sn in combination with a mass extrapolation for $^{101}$Sn, this irregularity was removed, recovering smooth trends not only in the tin chain but also in the tellurium and xenon chains. 

The mass uncertainty of five other nuclides near the proton drip line, linked to the tin masses through $\alpha$-decays, was thus improved.
This result highlights the necessity for precision mass spectrometry of short-lived radioactive nuclei, since inaccurate beta-decay measurements with significant measurement uncertainties --- in the order of a few hundreds of $\si{\kilo\electronvolt}$ --- can produce misleading trends of the mass filters.
%that hinder the accurate investigation of nuclear structure. 

In the course of submission of this article, we were made aware of a recent preprint~\cite{ireland2024highprecisionmassmeasurement103sn} reporting a Penning-trap mass measurement value for $^{103}$Sn with a measured mass excess of $\SI{-67125.9\pm 3.7}{\kilo\electronvolt}$.
The authors of Ref.~\cite{ireland2024highprecisionmassmeasurement103sn} draw a similar conclusion considering the smoothness of the mass surface in the tin isotopic chain near ${N=53}$.

\section{\label{sec:Acknowledgements}Acknowledgements}
%The authors would like to thank the ISOLDE-RILIS team for providing support for the operation of the resonance ionization laser ion source throughout the experiment, in particular K. Chrysalidis for the laser setup, operation, and calculations for the thermal state population distribution of the tin atoms in the ion source. 
The authors gratefully acknowledge technical support from the ISOLDE operations team, the CERN SY-STI-RBS team, and fruitful discussions with U.~Köster.
They further acknowledge the support of the German Max Planck Society, the French Institut National de Physique Nucléaire et de Physique des Particules (IN2P3), the European Research Council (ERC) under the European Union’s Horizon 2020 research and innovation programme (Grant Agreements No.~682841 ‘ASTRUm’, 654002 ‘ENSAR2’, 101020842 ‘EUSTRONG’, and 861198 ‘LISA’), as well as the German Federal Ministry of Education and Research (BMBF; Grants No.~05P18HGCIA, 05P21HGCI1, and 05P21RDFNB).
L.N. acknowledges support from the Wolfgang Gentner Programme of the German Federal Ministry of Education and Research (Grant No.~13E18CHA).

The experiment was conducted by M.A.-K., M.Au, D.A., C.B., K.C., P.F., R.H., C.K., D. Lu., D.La., B.M., M.M\"{u}., M.Mu., S.N., L.N., Ch.S., and F.W.
Resources and supervision were provided by K.B. and L.S.  
The manuscript was prepared by L.N., M.A.-K., K.C., and S.N.
All authors contributed to the editing of the manuscript.

% \nocite{*}

\bibliography{bib.bib}% Produces the bibliography via BibTeX.

% \appendix

% \section{Appendixes}

% \section{A little more on appendixes}

\end{document}